\documentclass[12pt,reqno]{article}
\usepackage{fullpage, amsmath,amssymb}
\usepackage[pdftex]{graphicx}

\title{Protein Folding: A New Geometric Analysis}
\author{Walter A. Simmons\\
Department of Physics and Astronomy\\
University of Hawaii at Manoa\\
\bigskip
Honolulu, HI 96822\\
Joel L. Weiner\\
Department of Mathematics\\
University of Hawaii at Manoa\\
Honolulu, HI 96822}

\date{ 7/23/2008} 

\begin{document}

\maketitle

\newpage

\begin{abstract}
A geometric analysis of protein folding, which compliments many of the models in the literature, is presented.  We examine the process from unfolded strand to the point where the strand becomes self-interacting.  A central question is how it is possible that so many initial configurations proceed to fold to a unique final configuration.  We put energy and dynamical considerations temporarily aside and focus upon the geometry alone.    
We parameterize the structure of an idealized protein using the concept of a ribbon from differential geometry.  The deformation of the ribbon is described by introducing a generic twisting Ansatz.  The folding process in this picture entails a change in shape guided by the local amino acid geometry.  The theory is reparamaterization invariant from the start, so the final shape is independent of folding time.  We develop differential equations for the changing shape.
For some parameter ranges, a sine-Gordon torsion soliton is found.  This purely geometric waveform, has properties similar to dynamical solitons.  Namely:  A threshold distortion of the molecule is required to initiate the soliton, after which, small additional distortions do not change the waveform. 
In this analysis, the soliton twists the molecule until bonds form.
The analysis reveals a quantitative relationship between the geometry of the amino acids and the folded form. 
\end{abstract}

\section{Introduction}

In the more than half century, since it was established that the amino acid sequence of a protein molecule determines the unique folded configuration [1] and that unfolded proteins re-fold from some range of initial conditions to the same end state, a wide range of physics and geometry based models have been intensively studied.  Recent reviews can be found in [2],  [3], and [4]  and there are many books that deal with the subject [5], [6], [7].  
The fact that proteins fold spontaneously from a range of initial configurations to a unique end state, in spite of the small energy available is astonishing.  In particular, there are many forces involved [2] and each of these is related to a change in shape through some non-linear, non-local, (and possibly temperature dependent) material response tensor. 

Particularly challenging to our understanding is the initial phase, which ends with the molecule becoming self-interacting.  The insensitivity to variation of initial conditions, immunity to noise, and to ambient conditions, leads us to conjecture that there is some essentially geometric aspect of folding that guides this initial stage.
In this paper, we put the dynamics, such as [8], and energy landscapes [3] temporarily aside and analyze the geometry of folding using differential geometry, which is the natural mathematical language for describing shapes and changes of shape.  Our analysis leads to a set of differential equations which are potentially useful to model-builders.  For a limited range of parameters, we solve the equations analytically and find that a torsion-wave soliton emerges.  This soliton lives in a space of possible molecular shapes, wherein it describes a twisting deformation, which ultimately stops when the molecule becomes self-interacting.  This soliton is different from the various dynamical solitons that have appeared in the literature for some time, e.g. [9], but it has similar characteristics; namely, a threshold for formation, a stability against noise, hierarchy of forms, and a non-linear superposition principle.

Our starting point is a construct known as a ribbon, which is a pair of writhing space curves.  One space curve, the base curve, aligns with an average backbone of the molecule.  The other curve, the neighboring curve, carries information about the local (amino acid) geometry, especially the location of the side-chains, and writhes about the base curve. (The ribbon has been previously applied to double-stranded DNA [10] and is compatible with the coils and kinks seen in protein molecules.)  We introduce a deformation Ansatz, which is a generic response to any torque. We have constructed the Ansatz so that the theory is reparameterization invariant; the folding is the same no matter how quickly or slowly it happens.
Differential geometry then leads us to a set of differential equations, which will be discussed in some generality elsewhere.  In this paper we focus upon a sub-set, which arises by making specific, empirically motivated, assumptions about the parameters, and which leads to the soliton.  
We discuss the solutions to the soliton and discuss how they can be accommodated to a structure which is segmented, not continuous.
One surprising property of the simplest, ÔantikinkÕ, solution is that the more planar  the initial shape of the of the unfolded molecule, the faster the molecule folds.  This property may not generalize to other solutions.

Before presenting our calculations in the next section, we conclude this section by remarking that it would be natural to combine our results with models based upon torsion angle [8] and/or energy landscape.  If the geometrical features described here also appear in models with dynamics, then a step will have been taken toward understanding insensitivity of folding to initial conditions, to noise, and to environmental factors.
We also remark that the reparameterization invariance  in our analysis are also encouraging.

\section{Ribbons and their deformations}

In this section we discuss ribbons and a certain kind of  deformation associated to them.  A ribbon can be  viewed as a space  curve with a  field of planes tangent to the curve.  It is reasonable for a discussion of ribbons to start with a discussion  of the differential geometry of space  curves.  Thus let  $\vec{x}: [0,L] \to \mathbb{E}^3$ be an embedded, i.e. one-to-one, curve in Euclidean  3-space given as a function of arc length $s$, where $L$ is the length of the curve $\vec{x}$.  We suppose that $\vec{x}$ has enough differentiability so that all that we discuss exists.  We may define the unit tangent vector field 
\[
\vec{e}_1 =  \vec{x}_s, 
\] 
where the subscript denotes differentiation with respect  to $s$.  If we assume that the curve $\vec{x}$ is non-degenerate, i.e., that $\vec{x}_s$ and $\vec{x}_{ss}$ are linearly independent at all points of the curve, then we can complete the vector field $\vec{e}_1$ to the Frenet frame $\vec{e}_1$, $\vec{e}_2$, $\vec{e}_3$, where $\vec{e}_2$ is the principal normal and $\vec{e}_3$ is the binormal.  This moving frame satisfies the well-known Frenet-Serret equations.
\begin{eqnarray*}
d\vec{e}_1 & = &\quad\qquad \kappa\vec{e}_2\, ds\\
d\vec{e}_2  & = &(-\kappa \vec{e}_1  \quad\qquad  + \tau \vec{e}_3) ds\\
d\vec{e}_3 & = &\ \qquad -\tau \vec{e}_2\, ds
\end{eqnarray*}
The functions $\kappa$ and $\tau$ give the curvature and torsion of the space curve $\vec{x}$, respectively.

For later use, we wish to point out that we may view $\vec{e}_1$ as a curve  taking values in $\mathbb{S}^2$,  the unit sphere  centered at the origin of $\mathbb{E}^3$.  If we represent $\vec{e}_1$ as a function of its arc length $\sigma$, then one may write  $\vec{e}_1:[0,K] \to \mathbb{S}^2$, where $K = \int_0^L \kappa\,  ds$ is the length of $\vec{e}_1$.  It  follows from the definition of the curvature $\kappa$  that
\begin{equation}\label{kappa}
\frac{d\sigma}{ds} = \kappa.
\end{equation}
  It is straightforward that the geodesic curvature $k$ of $\vec{e}_1$ is given by
\begin{equation}\label{k}
k = \frac{\tau}{\kappa}.
\end{equation}

To obtain a ribbon from a space curve we need to associate to the space  curve a field of planes tangent to the space  curve.  Since at each point of the space  curve, the vector $\vec{e}_1$ lies in the plane tangent at that point, the plane is completely determined by giving  a vector $\vec{\nu}$ that is perpendicular to $\vec{e}_1$ that lies in the tangent plane.   Necessarily $\vec{\nu}$ is in the plane spanned by $\vec{e}_2$ and $\vec{e}_3$.  Thus we introduce the function $\psi: [0,K]  \to \mathbb{R}$ by requiring that the following hold:
\begin{equation}\label{nu}
\vec{\nu} = \cos \psi \,\vec{e}_2 + \sin \psi\,\vec{e}_3.
\end{equation}

The space curve $\vec{x}$ we introduced represents the base curve of our model.   The neighboring curve in our model is represented by $\vec{x} + f \vec{e}_2$, where $f$ is a positive function of the arc length $s$.  Our primary interest in this section lies in how this ribbon deforms under a twisting operation on the base curve which depends upon the neighboring curve at every point.  More specifically we are interested in an essentially adiabatic, reparameterization invariant change in the shape of the base curve.  To this end, we parameterize the variation of the ribbon by means of a parameter $u$.  Thus all quantities under consideration become functions of $s$ and $u$.  For example, the torsion $\tau(s)$ becomes $\tau(s,u)$.  In this section, we examine the variation in the geometry of the ribbon  as  u ranges over some domain by means differential equations in the geometric invariants  mentioned  in the preceding  paragraphs. 

Our ansatz is the following  equation:
\[
\frac{\partial \vec{e}_1}{\partial u}(s,u) =  \gamma(u) f(s) \vec{\nu}(u,s).
\]

The coefficient $\gamma(u)$ is  a positive function chosen for later  convenience.  Moreover, it is a function of $u$ so that the form of the  equation is invariant under changes of the parameter $u$.  We consider any parameter which  is a continuously differentiable function of $u$ with positive derivative as  an  admissible parameter  for representing the variation.  Since $u$ will change over some interval, time dependence enters indirectly through this parameter.  We emphasize that in this section we are  studying changes in shape  under a twisting  deformation, not  dynamics.

Given what has just been said  we may as well choose a parameter $u$ for which $\gamma  = 1$. Thus we study the variation in the ribbon induced by the following differential equation:
\begin{equation}\label{physics}
\frac{\partial \vec{e}_1}{\partial u}(s,u) =  f(s) \vec{\nu}(s,u).
\end{equation}

We study this variation under the following assumptions:  If we view the base curve as a polygon with atoms at the vertices, then the lengths of the segments and the angles  between successive segments will remain constant under the variation.  In our model which is differentiable this corresponds to the following:
\begin{enumerate}
\item The element of arc  length $ds$ is invariant during the variation.
\item  The curvature $\kappa$ is invariant  during the variation.
\end{enumerate}

It follows from these assumptions and equation (\ref{kappa}) that the element  of arc $d\sigma$ remains unchanged  during the variation.

It is well-known that the shape of the base  curve $\vec{x}$ is completely determined by $\kappa$ and $\tau$ given  as functions of the arc length  $s$.  Since $\kappa$ does not depend on $u$, we regard it as a known function.  Thus our goal is to  determine how $\tau$ depends on $u$.  Since  $\tau = k \kappa$, we can just as well determine how $k$ depends on  $u$.
Finally,  our ansatz can be viewed as defining a variation of the curve $\vec{e}_1$.  Thus we transfer the variational problem to sphere $\mathbb{S}^2$,  and study how $\vec{e}_1$ varies under our ansatz which amounts to studying how $k$ varies under our ansatz.

In what follows we use the ``method of moving frames" and differential forms to make our calculations.  The  reader may want use a text  by [11] as a reference for what is done  below.

We summarize all we know about $\vec{e}_1$ and the variation in the following equations, where $v$ represents a quantity to be determined.  These equations follow from the Frenet-Serret equations and equations (\ref{kappa}), (\ref{k}), (\ref{nu}) and (\ref{physics}).
\begin{eqnarray}\label{variational eq1}
d\vec{e}_1 &  = &\qquad  \vec{e}_2 \,  d\sigma \qquad+ (f\cos \psi \,\vec{e}_2 + f\sin \psi \,\vec{e}_3 )\,  du\\
\label{variational eq2}
d\vec{e}_2  & =  & (- \vec{e}_1 + k\, \vec{e}_3)\,  d\sigma   +(  -f \cos \psi \,\vec{e}_1  + v\,\vec{e}_3  )\, du\\
\label{variational eq3}
d\vec{e}_3  & =  &\quad  - k\,\vec{e}_2\,  d\sigma  \qquad +  ( -f\sin\psi\,\vec{e}_1 - v\,\vec{e}_2)\,  du
\end{eqnarray}

We compute the exterior derivatives of the above equations  and  use the fact that $d^2 \vec{e}_i  = 0$.  From  the  $\vec{e}_2$ and $\vec{e}_3$ components of $d^2\vec{e}_1$ and $\vec{e}_3$ component of 
$d^2\vec{e}_2$, we get the following  equations.
\begin{eqnarray}\label{kis}
k & = & -\psi_{\sigma} + \frac{f_{\sigma}}{f} \cot \psi\\
\label{wis}
v & = & f_{\sigma} \csc \psi\\
\label{kuis}
0 & = &-k_u + w_{\sigma} + f \sin \psi 
\end{eqnarray} 

Using equations (\ref{kis}) and (\ref{wis}), we substitute for $k$ and $w$ in equation (\ref{kuis}) to obtain the following second order p.d.e. in $\psi$.
\begin{equation}\label{pde}
[\psi_{\sigma} - \frac{f_{\sigma}}{f} \cot \psi]_u + [f_{\sigma} \csc \psi]_\sigma + f \sin \psi = 0
\end{equation}

If we make the further assumption that $f$ is constant, this equation becomes
\begin{equation}\label{sinG}
\psi_{\sigma u} + f \sin \psi = 0,
\end{equation}
the sine-Gordon equation.

We  consider the implications of this equation given that it has soliton solutions.  If the base  curve is initially fairly planar, i.e.,  its torsion (or equivalently $k$) is not too far from zero  along its entire length, we consider how  a solition might explain the folding which is always observed.  Note, for  later use,  if $k$ is close to zero over the entire length of the base curve, it follows from equation (\ref{kis}) (assuming $f$ is constant), that $\psi$ is close to being constant.

We base  our arguments on   antikinks which for us take the form
\begin{equation}\label{antikink}
\psi(\sigma,u) = 4\tan^{-1}\left[\exp\left(\sqrt{f}(au -\frac{\sigma}{a} + b)\right)\right],
\end{equation}
where $a > 0$ and $b$ are constants.

We need to recall that the partial differential equation (\ref{pde}), and hence (\ref{sinG}), is defined on the  domain $[0,K]\times [0, U]$,  where $U$ may be some positive real or  $\infty$.  Thus one must consider either equation as  part of an initial value problem on that  domain.  Since the curves $\sigma =  {\rm constant}$  and $u = {\rm constant}$ are the characteristic curves of either  partial differential equation, we need  to consider our problem as a characteristic initial value problem.   Thus it is natural to suppose  that the values of $\psi$  are given on the curves $u = 0$ and $\sigma  = 0$.

Then we must accept as known that  
\[
\psi(\sigma, 0) = 4\tan^{-1}\left[\exp\left(\sqrt{f}( -\frac{\sigma}{a} + b)\right)\right].
\]
Given that initially we assume the base curve is fairly planar, the function   $\psi$ is close to being constant function on $[0,K]$.  We choose the constant $b$  so that 
$4\tan^{-1}\left[\exp\left(\sqrt{f}(  b)\right)\right]$ approximates that constant value and $a$ very large so that $4\tan^{-1}\left[\exp\left(\sqrt{f}( -\frac{\sigma}{a} + b)\right)\right]$ approximates that constant value on the interval $[0,K]$.  

We must also accept as known that
\[
\psi(0,u) = 4\tan^{-1}\left[\exp\left(\sqrt{f}(au+ b)\right)\right].
\]
If  a vibration of the left end point of the base curve be can represented by this function, then the antikink given by equation (\ref{antikink}) describes  the subsequent motion of the base curve in the following fashion.  As the $u$ increases in value, the antikink moves along the base curve and simultaneously, due to equation (\ref{kis}) a ``bump" of geodesic curvature, which  corresponds to a ``bump" of torsion moves along the base curve.  The effect of this  ``bump" of torsion is to twist the base curve into a position where  bonds are sure  to be  formed.   At this point, our model  is no longer viable and the ``bump" leaves in its wake the twisted form  of  the base chain of the protein.  

Should the initial values of $\psi$ be different and the vibrations of the left end point be of a different form as  well,  one can presume that there are other solitons  which satisfy these initial conditions and thus ultimately produce a twist that moves along the base curve leading to the formation a stable twisted molecule.

The process just described  for antikinks can, in fact, lead to similar conclusions if one assumes that $f$  is a piecewise constant function, rather then a constant function on $[0,K]$.  Let's suppose that  $f$ takes the value $f_1$ and $f_2$ on  the subintervals $[0,\sigma_1]$ and $[\sigma_1,\sigma_2]$ of $[0,K]$, respectively.  To construct a continuous, piecewise differentiable  solution  of equation (\ref{sinG}) we need the following to be true, for all $u$ in $[0,U]$:
\[
\sqrt{f_1}\left(a_1u  - \frac{\sigma_1}{a_1} + b_1\right) = \sqrt{f_2}\left(a_2u  - \frac{\sigma_1}{a_2} + b_2\right)
\]
If we regard $a_1$ and $b_1$ as known, we can clearly choose $a_2$ and $b_2$ for this to  be true.  Thus if again assume that $\psi$  is fairly constant on  $[0,K]$,   one can  still find solutions of equation (\ref{sinG}) that give rise to a moving ``bump" of torsion along the base curve. 

Our parameter  $u$ does not represent time but must be a monotonically increasing function of time.  Even though we are not dealing with dynamics, if we want to bring time into our considerations then our ansatz becomes
\begin{equation*}
\frac{\partial \vec{e}_1}{\partial t}(s,t) =  \gamma(t)f(s) \vec{\nu}(s,t),
\end{equation*}
for some positive real-valued function  $\gamma(t)$.
One easily finds that  equation (\ref{pde}) becomes
\begin{equation*}
[\psi_{\sigma} - \frac{f_{\sigma}}{f} \cot \psi]_t + \gamma [f_{\sigma} \csc \psi]_\sigma + \gamma f \sin \psi = 0,
\end{equation*}
and the sine-Gordon equation takes the form
\begin{equation*}
\psi_{\sigma t} + \gamma f \sin \psi = 0.
\end{equation*}
  The formula for an antikink becomes
\begin{equation*}
\psi(\sigma,t) = 4\tan^{-1}\left[\exp\left(\sqrt{f}(ag(t) -\frac{\sigma}{a} + b)\right)\right],
\end{equation*}
where $\frac{dg}{dt} = \gamma$ and $g(0) = 0$.

If we suppose $g$ is fairly constant and depends primarily the upon medium in which the protein is found (as opposed  to depending  upon the protein, itself) then we can argue as follows.  The more planar the initial shape of the base curve, the more closely  to being constant the initial values of $\psi$ are.  Hence,  the larger the value of $a$ must  be so that the antikink  approximates well those initial values.  However, the larger the value of $a$, the faster the soliton moves along the base curve, and hence the faster the ``bump" of torsion moves along the base curve and consequently the faster the formation of the twisted molecule.  

\section{Conclusions}

In this paper we have presented the results of a purely geometric analysis of the protein folding process.  Our most important results are as follows.

\begin{enumerate}
\item[i.] We parameterized a course-grained model of a protein molecule, which quantitatively describes the shape only of the molecule; the description is independent of position and orientation.  The parameterization includes the backbone and a distillation of the geometry of the side-chains that we refer to as the neighboring curve.  We call this model a ribbon.

This parameterization is particularly useful for geometrical and structural studies, providing that the coarse-graining is appropriate.  The ribbon is introduced to allow for analytic studies using differential geometry.
\item[ii.] We extended this parameterization to include changes of shape.  This parameterization that, again, depends only upon shape, not upon position, orientation, or space-time motion, defines a curve, i.e., a trajectory, of possible protein shapes.  We constrained the possible shapes in some ways that are appropriate by fixing length and bending. Changes in the twisting of the protein shapes is allowed and reflects the presence of amino acids through the idealized neighboring curve acting on the backbone.  

A molecule in this description follows some trajectory from unfolded to partly folded, where chemical bonds form. 

Our formulation is reparameterization invariant from the outset, therefore the folding geometry in independent of wall-clock time.

\item[iii.] Most importantly, we used the above parameterization to study possible trajectories in the case of an ad hoc by not unreasonable constraint on the shape.  We found a trajectory associated to a soliton solution of the sine-Gordon equation, producing what one might call a torsion soliton.  The soliton produces a torsional distortion of the molecule.  The distortion of the molecule is fixed, because of bond formation, during propagation of the soliton.

This soliton, which arises from geometric relationships within the folding molecule, is geometrical and thus different from dynamical solitons that are well known in protein science.  However, it has similar properties.
\begin{enumerate}
\item[a.] Its stability, i.e., the fact that propagates without continuous input of energy, is indifferent to scattering, temperature or forces that may vary from cell to cell,  may explain how it is possible that the folding is unique in spite of the variety of forces, response tensors, and environmental conditions involved.
\item [b.] If this soliton occurs in nature, it may also explain the other puzzles that were raised in the Introduction; a threshold distortion of the molecule is required to establish the soliton, but the sensitivity to initial conditions is minimal.
\item[c.] The soliton describes a self-focusing torsional wave.  Since energy was temporarily put aside at the outset, the energetics of the soliton remains unspecified here.  Obviously, the soliton will be relevant only if it is energetically allowed, but the fact that it arises from geometry alone suggests that a relatively flat energy landscape might suffice.  
\end{enumerate}

\end{enumerate}

\bigskip

\begin{center}
\Large
\textbf{List of References}
\normalsize
\end{center}
\bigskip

\begin{enumerate}

\item  Anfinsen CB (1973) Principles that govern the Folding of Protein Chains. \emph{Science} 181: 223-230.

\item Dill KA, Ozkan B,  Weikl TR (2008) The Protein Folding Problem. \emph{Annu Rev Biophys}  37: 289-316.

\item Onuchic JN, Luthey-Schulten Z, Wolynes G (1997) Theory of Protein Folding: The Energy Landscape Perspective. \emph{Annu Rev Phys Chem} 48: 545-600.

\item Finkelstein AV, Galzitskaya OV (2004) Physics of Protein Folding. \emph{Physics of Life Reviews} 1:  23-56. 

\item Fersht A (1991) \emph{Structure and Mechanism in Protein Science}, (W.H. Freeman \& Co., New York).

\item Finkelstein AV, Ptitsyn  OB (2002) \emph{Protein  Physics}, (Academic Press, London).

\item Wales DJ (2003) \emph{Energy Landscapes}, (Cambridge, New York).

\item  Guntert P, Mumenthaler C, Wuethrich K (1997) Torsion Angle Dynamics for NMR Structure Calculation with the New Program DYANA. \emph{J Mol Biol} 273: 283-298.

\item Davydov AS (1971) \emph{Theory of Molecular Excitations}, (Plenum Press, New York-London).

\item  White J H,  Bauer WR (1986) Calculation of the Twist and the Writhe  for representative models of DNA. \emph{J Mol Biol} 189: 329-341.

\item O'Neill B  (1997) \emph{Elementary Differential Geometry, Second  Edition},  (Academic Press, San Diego,  London).

\end{enumerate}

\end{document}